\documentclass[aps,preprint,showpacs,groupedaddress]{revtex4-1}
\usepackage{amssymb}
\usepackage{mathrsfs}
\usepackage{axodraw4j}
\usepackage{pstricks}
\usepackage{color}
\usepackage{feynMF}
\usepackage{graphicx}
\usepackage{axodraw}

\begin{document}


\title{\bf Mass Renormalization in Lorentz-violating Scalar Field Theory}



\author{Paulo R. S. Carvalho}
\email{prscarvalho@ufpi.edu.br}
\affiliation{\it Departamento de F\'\i sica, Universidade Federal do Piau\'\i, 64049-550, Teresina, PI, Brazil}




\begin{abstract}
In this work we evaluate the $\gamma_{m}$ function corresponding to mass renormalization for $O(N)$ scalar field theory with Lorentz violation. We calculate this function up to two-loop order for a theory renormalized utilizing the counterterm method in the minimal subtraction scheme with Feynman diagrams regularized using dimensional regularization.
\end{abstract}


\maketitle


\section{Introduction}\label{Introduction}

\par The standard model (SM) of elementary particles and fields was built along many years by theoretical, phenomenological and experimental arguments. Some of the theoretical arguments emerged following symmetry ideas and the resultant SM as we know today is invariant under Lorentz, gauge and CPT symmetries. On the other hand, possible Lorentz-violating (LV) extensions of the SM, standard model extensions (SME), have, apparently, no troubles both theoretically and experimentally \cite{PhysRevD.58.116002, Koste} and many theories exhibiting Lorentz violation were proposed in recent years. We can mention the cases of the SM Electroweak sector \cite{PhysRevD.79.125019}, Quantum Chromodynamics \cite{PhysRevD.77.085006}, Quantum Electrodynamics \cite{PhysRevD.65.056006}, a pure Yang-Mills theory \cite{PhysRevD.75.105002} and other theories \cite{PhysRevD.39.683, Altschul2006679, PhysRevD.84.065030, PhysRevD.86.125015, Boldo2010112, PhysRevLett.102.251601, deBerredoPeixoto2006153}. 

\par Even at low energies, if we want to study phase transitions and critical phenomena using perturbative field-theoretical methods, for example for Lifshitz points, we have to renormalize LV theories as well \cite{Carvalho2010151, Carvalho2009178, PhysRevB.72.224432, PhysRevB.67.104415}. In this case, the breaking of Lorentz symmetry is easily seen through the free propagator of the theory whose dispersion relation is not quadratic. For the so called anisotropic $m$-axial Lifshitz points, the inverse of the massless Euclidean free propagator is $q^{2}+(k^{2})^2$ where the quadratic momentum $q$ is ($d-m$)-dimensional while the quartic momentum $k$ has $m$ components and the Lorentz-symmetry breaking occur in a $m$-dimensional subspace. When we are faced with the isotropic $m$-axial situation, the referred propagator assumes the form $(k^{2})^{2}$ and we have a non quadratic dispersion relation once again. So in high and low energy physics we can see that theories admitting the violation of Lorentz symmetry is an important issue nowadays.

\par We know that quantum field theories are plagued by divergences. Unfortunately divergent theories are senseless because its physical quantities are unbounded \cite{PhysRev.85.631}. However, any acceptable theory designed to describe physical phenomena with many particles and fields must be finite. The cure for these infinities was provided by the renormalization group technique  \cite{PhysRevB.4.3174}. The renormalization group is a tool for handling divergences and hence extract the information necessary to obtain reliable physical results.

\par An attempt of having a finite quantum field theory for the SME could begin with the renormalization of its Higgs sector, which involves a scalar field. The self-interacting scalar field theory has three nonrenormalized parameters: the field $\phi_{B}$, its mass $m_{B}$ and the self interacting coupling constant $\lambda_{B}$. If we desire a full renormalized quantum field-theoretic version of this theory we have to renormalize all the three physical quantities just mentioned. There are many ways to achieve such a goal. The method used for this purpose is the counterterm method in the minimal subtraction scheme (MS) \cite{'tHooft1972189} using dimensional regularization (DR) \cite{Bollini19725669, Bollini1972566} for computing the Feynman diagrams . In this scheme we renormalize the vertex functions leaving the external momenta arbitrary. Indeed, in this scheme, the theory is renormalized at nonzero external momenta denoting its generality and elegance. Besides all, it preserves gauge symmetry and has been highly suitable for treating gauge theories. The renormalized two-point function is written as a Taylor expansion in powers of external momenta $P^{2}$ and renormalized mass $m^{2}$. The coefficient proportional to $P^{2}$ is used to renormalize the field and calculate the corresponding Wilson $\gamma$ function. The term proportional to $m^{2}$ permits us obtain the other Wilson $\gamma_{m}$ function associated to mass renormalization. When we renormalize the four-point function we can compute the $\beta$ function of the theory. The existence of three parameters to be renormalized gives rise to three conditions which are satisfied by theory when the external momenta are fixed at zero. These conditions are called renormalization conditions, two of which are imposed for the two-point function and the remaining condition for the four-point function. We can use power counting arguments to understand why we have to have three renormalization conditions: the former has two divergences, one squared and one logarithmic while the only divergence to be absorbed in the latter function is logarithmic. Thus, if we want a renormalized theory, we have to renormalize these two functions. 

\par Moreover, the Higgs sector would require the knowledge of renormalization properties of LV scalar fields. With this intention, the scalar field theory with Lorentz violation was renormalized and the quantum corrections to $\gamma$ function at two-loop level and $\beta$ function up to next-to-leading order were calculated \cite{PhysRevD.84.065030}. All calculations for obtaining the desired functions mentioned above were performed explicitly for second order in a set of small constant parameters $K_{\mu\nu}$ to be defined below. In a proof by induction, the theory was renormalized up to a finite general order in these parameters. By adding the contributions for all orders up to infinity in $K_{\mu\nu}$, the problem was solved exactly for these constants. Then, the factor
\begin{eqnarray}
\Pi = 1 - \frac{1}{2}K_{\mu\nu}\eta^{\mu\nu} + \frac{1}{8}K_{\mu\nu}K_{\rho\sigma}\eta^{\{\mu\nu}\eta^{\rho\sigma\}} + ...
\end{eqnarray}
emerged from the exact solution where $\eta^{\mu\nu}$ is the four-dimensional Minkowski metric tensor and $\eta^{\{\mu\nu}\eta^{\rho\sigma\}} \equiv \eta^{\mu\nu}\eta^{\rho\sigma} + \eta^{\mu\rho}\eta^{\nu\sigma} + \eta^{\mu\sigma}\eta^{\nu\rho}$. This factor shows how the LV theory is modified by the symmetry breaking. However, the mass renormalization was not discussed. The mass renormalization can be viewed otherwise as a composite operator renormalization problem which is related to scattering amplitudes and thus represents a very important issue \cite{doi:10.1142/S0217732307020981}. Approaching the mass renormalization to next-to-leading order is the aim of this work and attains the task of obtaining a full renormalization of the theory. 

\par In this Letter we will show in Sec. \ref{O(N) scalar field theory with Lorentz violation} the renormalization of the $O(N)$ scalar field theory with Lorentz violation in the counterterm method in MS. In the Sec. \ref{Mass renormalization} we will discuss the evaluation of the renormalization constant for mass renormalization and the respective Wilson function. We will finalize the Letter in Sec. \ref{Conclusions} with our conclusions.

\section{O(N) scalar field theory with Lorentz violation}\label{O(N) scalar field theory with Lorentz violation}
\subsection{Bare theory}\label{Bare theory}
\par The $O(N)$ LV self-interacting massive scalar field theory has the unrenormalized or bare Lagrangian \cite{PhysRevD.84.065030} 
\begin{eqnarray}
\mathscr{L} = \frac{1}{2}\partial^{\mu}\phi_{B}\partial_{\mu}\phi_{B} + \frac{1}{2}K_{\mu\nu}\partial^{\mu}\phi_{B}\partial^{\nu}\phi_{B} + \frac{1}{2}m_{B}^{2}\phi_{B}^{2} + \frac{\lambda_{B}}{4!}\phi_{B}^{4}
\end{eqnarray}  
where the bare field $\phi_{B}$ is a vector field with $N$ components in a $d$-dimensional Euclidean space and $m_{B}$ and $\lambda_{B}$ are the bare mass and bare coupling constant, respectively. The dimensionless symmetric ($K_{\mu\nu} = K_{\nu\mu}$) constant coefficients $|K_{\mu\nu}|\ll 1$ are the same for all $N$ components of the vector field. They are very small in magnitude and provide a slight violation of Lorentz symmetry. The $O(N)$ internal symmetry implies that $\phi_{}^{2} = \phi_{1}^{2} + ... + \phi_{N}^{2}$ and $\phi_{}^{4} = (\phi_{1}^{2} + ... + \phi_{N}^{2})^{2}$. This Lagrangian is a modified version of the usual or Lorentz-invariant (LI) scalar field theory with a different inverse free propagator in momentum space given by $q^{2} + K_{\mu\nu}q^{\mu}q^{\nu} + m_{B}^{2}$. Thus to get a renormalized theory, we have to renormalize the bare two- and four-point vertex functions $\overline{\Gamma}_{B}^{(2)}$ and $\overline{\Gamma}_{B}^{(4)}$. This is what we will do in the next section.

\subsection{Method of counterterms for renormalized theory}\label{Method of counterterms for the renormalized theory}
\par As seen in Sec. \ref{Introduction}, the renormalized two-point function can be written as a Taylor expansion of external momentum $P^{2}$ and renormalized mass $m^{2}$ as well \cite{Kleinert}. In the reference \cite{PhysRevD.84.065030}, the term proportional to $P^{2}$ of this function was used for calculating the $\gamma$ function corresponding to field renormalization. The same authors renormalized the four-point function and obtained the $\beta$ function associated to coupling constant renormalization. If we want to renormalize the mass, we have to calculate the term proportional to $m^{2}$ of the two-point function. This is what will be done in this section following the steps in the reference \cite{Kleinert}.

\par The renormalized theory is not attained by using renormalization conditions, although the final renormalized theory satisfies these conditions at vanishing external momenta \cite{Kleinert}, and is otherwise obtained adding to its initial Lagrangian a few terms. These terms will generate additional diagrams called counterterm diagrams which will render the theory finite. The corresponding renormalized $n$-point functions with $n\geq 1$ satisfy the Callan-Symanzik equation 
\begin{eqnarray}
\left[\mu\frac{\partial}{\partial\mu} + \beta(g)\frac{\partial}{\partial g} - n\gamma(g) + \gamma_{m}(g)m\frac{\partial}{\partial m}\right]\overline{\Gamma}^{(n)}(k_{1},...,k_{n};m,g,\mu) = 0
\end{eqnarray}
where the $\beta$ and $\gamma$ functions were evaluated earlier \cite{PhysRevD.84.065030} and
\begin{eqnarray}
\gamma_{m} = \frac{\mu}{m}\frac{\partial m}{\partial\mu}\Bigg\vert_{B}
\end{eqnarray}
where $|_{B}$ means that the respective parameters must be that of bare theory. In particular, the renormalized two-point function acquires the form
\begin{eqnarray}
\overline{\Gamma}^{(2)} & = & \quad
\begin{picture}(33,2) (113,-160)
    \SetWidth{1.0}
    \SetColor{Black}
    \Line(114,-159)(145,-159)
\end{picture}^{-1} \quad - \quad  
\begin{picture}(30,17) (212,-153)
    \SetWidth{0.9}
    \SetColor{Black}
    \Arc(227,-144)(7.071,262,622)
    \Line(213,-152)(241,-152)
    \Vertex(227,-151.5){1.5}
\end{picture} \quad - \quad
\begin{picture}(51,13) (111,-204)
    \SetWidth{1.0}
    \SetColor{Black}
    \Line(118,-198)(157,-198)
    \SetWidth{1.0}
    \Line(133,-192)(142,-203)
    \Line(142,-192)(133,-203)
\end{picture}\quad - \quad 
\begin{picture}(68,14) (65,-167)
    \SetWidth{1.0}
    \SetColor{Black}
    \Line(84,-160)(115,-160)
    \Arc(100,-160)(3,270,630)
\end{picture}  \nonumber \\ & & \quad - \quad 
\begin{picture}(28,30) (204,-138)
    \SetWidth{0.9}
    \SetColor{Black}
    \Arc(218,-129)(6.083,261,621)
    \Line(205,-136)(231,-136)
    \Vertex(218,-135.5){1.5}
    \Arc(218,-116)(6.083,261,621)
    \Vertex(218,-122){1.5}
\end{picture} \quad - \quad
\begin{picture}(29,18) (152,-182)
    \SetWidth{1.0}
    \SetColor{Black}
    \Arc(166,-173)(8.246,256,616)
    \Line(152,-173)(180,-173)
    \Vertex(158,-173){1.5}
    \Vertex(174,-173){1.5}
\end{picture} \quad - \quad 
\begin{picture}(16,21) (216,-193)
    \SetWidth{1.0}
    \SetColor{Black}
    \Arc(224,-182)(7,270,630)
    \Line(215,-190)(233,-190)
    \Vertex(224,-189.5){1.5}
    \Line(221,-172)(227,-179)
    \Line(227,-172)(221,-179)
\end{picture} \quad - \quad 
\begin{picture}(30,17) (212,-153)
    \SetWidth{1.0}
    \SetColor{Black}
    \Arc(227,-144)(7.071,262,622)
    \Line(213,-152)(241,-152)
    \Vertex(227,-151){3}
\end{picture}
\end{eqnarray}
where the third, fourth and last two diagrams are couterterms diagrams. All these diagrams were regularized using DR 
\begin{eqnarray}
\int \frac{d^{d}q}{(2\pi)^{d}} \frac{1}{(q^{2} + 2pq + M^{2})^{\alpha}} = \hat{S}_{d}\frac{1}{2}\frac{\Gamma(d/2)}{\Gamma(\alpha)}\frac{\Gamma(\alpha - d/2)}{(M^{2} - p^{2})^{\alpha - d/2}}    
\end{eqnarray}
with $\hat{S}_{d}=S_{d}/(2\pi)^{d}=2/(4\pi)^{d/2}\Gamma{(d/2)}$ where $S_{d}=2\pi^{d/2}/\Gamma(d/2)$ is the surface area of a unit $d$-dimensional sphere and has the finite value $\hat{S}_{4}=2/(4\pi)^{2}$ in four-dimensional space. Therefore each loop integration contributes with a factor $\hat{S}_{4}$ in four dimensions. The integral written in the form above was used for convenience and avoids the appearance of Euler-Mascheroni constants in the middle of calculations \cite{Amit}. When we do not use this form, all these constants appear and have to cancel precisely in the final renormalized theory. In the next section the Feynman diagrams will be calculated in a $\varepsilon$-expansion where $\varepsilon = 4 - d$.

\section{Mass renormalization}\label{Mass renormalization}
\subsection{Evaluation of $\gamma_{m}$ function}
\par Before regularizing the Feynman diagrams we have to write the renormalized free propagator and its powers in terms of the small parameters $K_{\mu\nu}$ as
\begin{eqnarray}\label{expansion}
\frac{1}{(q^{2} + K_{\mu\nu}q^{\mu}q^{\nu} + m^{2})^{n}} = \frac{1}{(q^{2} + m^{2})^{n}} \left[ 1 - n\frac{K_{\mu\nu}q^{\mu}q^{\nu}}{q^{2} + m^{2}} +  \frac{n(n+1)}{2!} \frac{K_{\mu\nu}K_{\rho\sigma}q^{\mu}q^{\nu}q^{\rho}q^{\sigma}}{(q^{2} + m^{2})^{2}} + ...\right]
\end{eqnarray}
where the diagrams were calculated up to $\mathcal{O}(K^{2}$) (see Appendix). In the process of integration, after the expansion in eq. (\ref{expansion}), we will have to calculate many integrals. Many of them are not independent and we can show that some integrals are the same by a simple change of variables of integration. This procedure reduces the number of integrals to be evaluated. As our calculations occur in Euclidean space, we have by a simple Wick rotation $\eta^{\mu\nu} \rightarrow \delta^{\mu\nu}$ and 
\begin{eqnarray}
\Pi \rightarrow 1 - \frac{1}{2}K_{\mu\nu}\delta^{\mu\nu} + \frac{1}{8}K_{\mu\nu}K_{\rho\sigma}\delta^{\{\mu\nu}\delta^{\rho\sigma\}} + ...
\end{eqnarray}
where $\delta^{\{\mu\nu}\delta^{\rho\sigma\}} \equiv \delta^{\mu\nu}\delta^{\rho\sigma} + \delta^{\mu\rho}\delta^{\nu\sigma} + \delta^{\mu\sigma}\delta^{\nu\rho}$.

\par The mass is renormalized by a renormalization constant given by
\begin{eqnarray}\label{Z_{m^{2}}}
Z_{m^{2}}(g,\varepsilon^{-1}) = 1 + \frac{1}{m^{2}} \Biggl[ \frac{1}{2} \mathcal{K}
\left( \begin{picture}(30,17) (212,-153) 
    \SetWidth{0.9}
    \SetColor{Black}
    \Arc(227,-144)(7.071,262,622)
    \Line(213,-152)(241,-152)
    \Vertex(227,-151.5){1.5}
  \end{picture} \right)S_{\scalebox{0.3}{\begin{picture}(30,17) (212,-153) 
    \SetWidth{0.9}
    \SetColor{Black}
    \Arc(227,-144)(7.071,262,622)
    \Line(213,-152)(241,-152)
    \Vertex(227,-151.5){1.5}
  \end{picture}}} +
  \frac{1}{4} \mathcal{K}
  \left( \parbox{8mm} {\begin{picture}(28,30) (204,-140)
    \SetWidth{0.9}
    \SetColor{Black}
    \Arc(218,-129)(6.083,261,621)
    \Line(205,-136)(231,-136)
    \Vertex(218,-135.5){1.5}
    \Arc(218,-116)(6.083,261,621)
    \Vertex(218,-122){1.5}
  \end{picture}} \hspace*{0.2cm}  \right)S_{\scalebox{0.3}{\begin{picture}(28,30) (204,-140)
    \SetWidth{0.9}
    \SetColor{Black}
    \Arc(218,-129)(6.083,261,621)
    \Line(205,-136)(231,-136)
    \Vertex(218,-135.5){1.5}
    \Arc(218,-116)(6.083,261,621)
    \Vertex(218,-122){1.5}
  \end{picture}}} \nonumber \\
   +  \frac{1}{2} \mathcal{K}
  \left( \parbox{8mm} {\begin{picture}(16,21) (216,-193)
    \SetWidth{1.0}
    \SetColor{Black}
    \Arc(224,-182)(7,270,630)
    \Line(215,-190)(233,-190)
    \Vertex(224,-189.5){1.5}
    \Line(221,-172)(227,-179)
    \Line(227,-172)(221,-179)
  \end{picture}} \right)S_{\scalebox{0.3}{\begin{picture}(16,21) (216,-193)
    \SetWidth{1.0}
    \SetColor{Black}
    \Arc(224,-182)(7,270,630)
    \Line(215,-190)(233,-190)
    \Vertex(224,-189.5){1.5}
    \Line(221,-172)(227,-179)
    \Line(227,-172)(221,-179)
  \end{picture}}} 
  + \frac{1}{2} \mathcal{K}
  \left( \begin{picture}(30,17) (212,-153)
    \SetWidth{1.0}
    \SetColor{Black}
    \Arc(227,-144)(7.071,262,622)
    \Line(213,-152)(241,-152)
    \Vertex(227,-151){3}
  \end{picture} \right)S_{\scalebox{0.3}{\begin{picture}(30,17) (212,-153)
    \SetWidth{1.0}
    \SetColor{Black}
    \Arc(227,-144)(7.071,262,622)
    \Line(213,-152)(241,-152)
    \Vertex(227,-151){3}
  \end{picture}}} +
  \frac{1}{6} \mathcal{K}
  \left( \parbox{8mm}{ \begin{picture}(29,18) (152,-182)
    \SetWidth{1.0}
    \SetColor{Black}
    \Arc(166,-173)(8.246,256,616)
    \Line(152,-173)(180,-173)
    \Vertex(158,-173){1.5}
    \Vertex(174,-173){1.5}
  \end{picture}}\hspace*{0.2cm} \right) \Biggr|_{p^2 = 0}S_{\scalebox{0.3}{\begin{picture}(29,18) (152,-182)
    \SetWidth{1.0}
    \SetColor{Black}
    \Arc(166,-173)(8.246,256,616)
    \Line(152,-173)(180,-173)
    \Vertex(158,-173){1.5}
    \Vertex(174,-173){1.5}
  \end{picture}}} \Biggr].
\end{eqnarray}
The renormalization constant $Z_{m^{2}}(g,\varepsilon^{-1})$, one of the constants which absorbs the divergences of the theory, is a function of renormalized dimensionless coupling constant $g = \lambda\mu^{-\varepsilon}$, where $\mu$ is an arbitrary mass parameter, and a Laurent expansion in $\varepsilon$. The operator $\mathcal{K}$ extracts only the divergent part of the diagrams. The factor $ 
S_{\scalebox{0.3}{\begin{picture}(30,17) (212,-153) 
    \SetWidth{0.9}
    \SetColor{Black}
    \Arc(227,-144)(7.071,262,622)
    \Line(213,-152)(241,-152)
    \Vertex(227,-151.5){1.5}
  \end{picture}}}$
is the symmetry factor for a scalar theory with $O(N)$ symmetry for the tadpole diagram and so on. The last diagram or commonly called the sunset diagram is the most hard to integrate. It will be solved in more details than the others which are simpler.  

\par The first diagram in eq. (\ref{Z_{m^{2}}}), the tadpole, was calculated early in \citep{PhysRevD.84.065030} and according to our conventions 
\begin{eqnarray}
\begin{picture}(30,17) (212,-153)
    \SetWidth{0.9}
    \SetColor{Black}
    \Arc(227,-144)(7.071,262,622)
    \Line(213,-152)(241,-152)
    \Vertex(227,-151.5){1.5}
  \end{picture}=
  -\lambda\int \frac{d^{d}q}{(2\pi)^{d}}\frac{1}{q^2 + K_{\mu\nu}q^{\mu}q^{\nu} + m^{2}} = \frac{2m^{2}g}{(4\pi)^{2}\varepsilon}\left[ 1 - \frac{1}{2}\varepsilon\ln\left(\frac{m^{2}}{4\pi\mu^{2}}\right)\right]\Pi.
\end{eqnarray}

\par The ``double scoop'' \citep{Ramond} or the second diagram has a tadpole on its expression, hence we have
\begin{eqnarray}
\begin{picture}(28,30) (204,-138)
    \SetWidth{0.8}
    \SetColor{Black}
    \Arc(218,-129)(6.083,261,621)
    \Line(205,-136)(231,-136)
    \Vertex(218,-135.5){1.5}
    \Arc(218,-116)(6.083,261,621)
    \Vertex(218,-122){1.5}
  \end{picture} = \lambda^{2}\int \frac{d^{d}q_{1}}{(2\pi)^{d}}\frac{d^{d}q_{2}}{(2\pi)^{d}}\frac{1}{(q_{1}^2 + K_{\mu\nu}q_{1}^{\mu}q_{1}^{\nu} + m^{2})^{2}}\frac{1}{q_{2}^2 + K_{\mu\nu}q_{2}^{\mu}q_{2}^{\nu} + m^{2}} = \nonumber \\ - \frac{4m^{2}g^{2}}{(4\pi)^{4}\varepsilon^{2}}\left[ 1 - \frac{1}{2}\varepsilon - \varepsilon\ln\left(\frac{m^{2}}{4\pi\mu^{2}}\right)\right]\Pi^{2}.
\end{eqnarray} 

\par The next two counterterm diagrams can be easily obtained by certain operations on tadpole and ``fish'' $\begin{picture}(30,24) (218,-161)
\SetWidth{0.8}
    \SetColor{Black}
    \Arc(233,-157)(8,252,612)
    \Line(241.5,-157)(246,-151)
    \Line(241.5,-157)(246,-163)
    \Line(224.5,-157)(220,-151)
    \Line(224.5,-157)(220,-163)
    \Vertex(224.5,-157){1.5}
    \Vertex(241.5,-157){1.5}
  \end{picture}$ 
diagrams calculated previously \cite{PhysRevD.84.065030}. The first and second counterterms then result
\begin{eqnarray}
\begin{picture}(16,21) (216,-193)
    \SetWidth{1.0}
    \SetColor{Black}
    \Arc(224,-182)(7,270,630)
    \Line(215,-190)(233,-190)
    \Vertex(224,-189.5){1.5}
    \Line(221,-172)(227,-179)
    \Line(227,-172)(221,-179)
  \end{picture} = \frac{2m^{2}g^{2}}{(4\pi)^{4}\varepsilon^{2}}\left[ 1 - \frac{1}{2}\varepsilon - \frac{1}{2} \varepsilon\ln\left(\frac{m^{2}}{4\pi\mu^{2}}\right)\right]\Pi^{2},
\end{eqnarray}
\begin{eqnarray}
\begin{picture}(30,17) (212,-153)
    \SetWidth{1.0}
    \SetColor{Black}
    \Arc(227,-144)(7.071,262,622)
    \Line(213,-152)(241,-152)
    \Vertex(227,-151){3}
  \end{picture} = \frac{6m^{2}g^{2}}{(4\pi)^{4}\varepsilon^{2}}\left[ 1 - \frac{1}{2} \varepsilon\ln\left(\frac{m^{2}}{4\pi\mu^{2}}\right)\right]\Pi^{2}.
\end{eqnarray}

\par Now we are left with the sunset diagram
\begin{eqnarray}\label{sunset}
\begin{picture}(29,18) (152,-177)
    \SetWidth{1.0}
    \SetColor{Black}
    \Arc(166,-173)(8.246,256,616)
    \Line(152,-173)(180,-173)
    \Vertex(158,-173){1.5}
    \Vertex(174,-173){1.5}
  \end{picture} = \lambda^{2}\int \frac{d^{d}q_{1}}{(2\pi)^{d}}\frac{d^{d}q_{2}}{(2\pi)^{d}}\frac{1}{q_{1}^2 + K_{\mu\nu}q_{1}^{\mu}q_{1}^{\nu} + m^{2}}\frac{1}{q_{2}^2 + K_{\mu\nu}q_{2}^{\mu}q_{2}^{\nu} + m^{2}} \nonumber \\ \times \frac{1}{(q_{1} + q_{2} + P)^2 + K_{\mu\nu}(q_{1} + q_{2} + P)^{\mu}(q_{1} + q_{2} + P)^{\nu} + m^{2}}.
\end{eqnarray}
This diagram can be written as two different integrals, each one being the coefficients of $P^{2}$ and $m^{2}$ powers, as seen in Sec. \ref{Introduction}, using an identity, the ``partial-$q$" $\partial q^{\mu}/\partial q^{\mu} = d$  \citep{'tHooft1972189}. We can introduce  
\begin{eqnarray}
1 = \frac{1}{2d}\left(\frac{\partial q_{1}^{\mu}}{\partial q_{1}^{\mu}} + \frac{\partial q_{2}^{\mu}}{\partial q_{2}^{\mu}}\right)
\end{eqnarray}
in eq. (\ref{sunset}) to obtain
\begin{eqnarray}\label{sunset_2}
\begin{picture}(29,18) (152,-177)
    \SetWidth{1.0}
    \SetColor{Black}
    \Arc(166,-173)(8.246,256,616)
    \Line(152,-173)(180,-173)
    \Vertex(158,-173){1.5}
    \Vertex(174,-173){1.5}
  \end{picture} = -\frac{\lambda^{2}}{d-3}[3m^{2}A(P) + B(P)]
\end{eqnarray}
where 
\begin{eqnarray}
A(P) = \int \frac{d^{d}q_{1}}{(2\pi)^{d}}\frac{d^{d}q_{2}}{(2\pi)^{d}}\frac{1}{q_{1}^2 + K_{\mu\nu}q_{1}^{\mu}q_{1}^{\nu} + m^{2}}\frac{1}{q_{2}^2 + K_{\mu\nu}q_{2}^{\mu}q_{2}^{\nu} + m^{2}} \nonumber \\ \times \frac{1}{[(q_{1} + q_{2} + P)^2 + K_{\mu\nu}(q_{1} + q_{2} + P)^{\mu}(q_{1} + q_{2} + P)^{\nu} + m^{2}]^{2}},
\end{eqnarray}
\begin{eqnarray}
B(P) = \int \frac{d^{d}q_{1}}{(2\pi)^{d}}\frac{d^{d}q_{2}}{(2\pi)^{d}}\frac{1}{q_{1}^2 + K_{\mu\nu}q_{1}^{\mu}q_{1}^{\nu} + m^{2}}\frac{1}{q_{2}^2 + K_{\mu\nu}q_{2}^{\mu}q_{2}^{\nu} + m^{2}} \nonumber \\ \times \frac{P(q_{1} + q_{2} + P) + K_{\mu\nu}P^{\mu}(q_{1} + q_{2} + P)^{\nu}}{[(q_{1} + q_{2} + P)^2 + K_{\mu\nu}(q_{1} + q_{2} + P)^{\mu}(q_{1} + q_{2} + P)^{\nu} + m^{2}]^{2}}.
\end{eqnarray}
The divergent part of the second integral is proportional to a power $P^{2}$ and can be used to renormalize the field and calculate the $\gamma$ function \cite{PhysRevD.84.065030}, namely $\gamma(g) = (N+2)g^{2}\Pi^{2}/36(4\pi)^{4}$. This function  will be necessary in the $\gamma_{m}$ function evaluation. The first integral is the sunset contribution to mass renormalization as seen symbolically in eq. (\ref{Z_{m^{2}}}) and (\ref{sunset_2}). For zero order in $\mathcal{K}$, introducing Feynman parameters \citep{Amit} and after momentum integration, this integral takes the form
\begin{eqnarray}
A^{(0)}(P) = \frac{1}{(4\pi)^{4}\varepsilon}(1 - \varepsilon)\int_{0}^{1}dx[x(1-x)]^{-\varepsilon/2}\int_{0}^{1}dyy^{\varepsilon/2 - 1}(1-y) \nonumber \\ \times \left\{\frac{y(1-y)P^{2}}{4\pi} + \left[1-y + \frac{y}{x(1-x)}\right]\frac{m^{2}}{4\pi}\right\}^{-\varepsilon}.
\end{eqnarray}  
This integral is singular for $y = 0$ when $\varepsilon = 0$. So its divergent contribution is obtained integrating $\{\}^{-\varepsilon}|_{y=0}$ instead of $\{\}^{-\varepsilon} $\cite{Kleinert, Amit}. Hence we have
\begin{eqnarray}
A^{(0)}(P) = \frac{2}{(4\pi)^{4}\varepsilon^{2}}\left[ 1 - \frac{1}{2}\varepsilon - \varepsilon\ln\left(\frac{m^{2}}{4\pi}\right)\right].
\end{eqnarray}     
If we include the $\mathcal{O}(K)$ and $\mathcal{O}(K^{2})$ orders for obtaining the divergent contributions of remaining parametric integrals, we finally find
\begin{eqnarray}
\left( \parbox{8mm}{ \begin{picture}(29,18) (152,-182)
    \SetWidth{1.0}
    \SetColor{Black}
    \Arc(166,-173)(8.246,256,616)
    \Line(152,-173)(180,-173)
    \Vertex(158,-173){1.5}
    \Vertex(174,-173){1.5}
  \end{picture}}\hspace*{0.2cm} \right) \Biggr|_{p^2 = 0} = - \frac{6m^{2}g^{2}}{(4\pi)^{4}\varepsilon^{2}}\left[ 1 + \frac{1}{2}\varepsilon - \varepsilon\ln\left(\frac{m^{2}}{4\pi\mu^{2}}\right)\right]\Pi^{2}.
\end{eqnarray}

\par Using these results, after cancellation of all mass logarithm terms which ensures the renormalizability of the theory, the mass renormalization constant is given by
\begin{eqnarray}
Z_{m^{2}} = 1 + \frac{N+2}{3(4\pi)^{2}\varepsilon}\Pi g + \left[\frac{(N+2)(N+5)}{9\varepsilon^{2}} - \frac{N+2}{6\varepsilon}\right]\frac{\Pi^{2}g^{2}}{(4\pi)^{4}}.
\end{eqnarray}

\par Now, calculating the $\gamma_{m}$ function we have 
\begin{eqnarray}\label{gamma m}
\gamma_{m}(g) = \frac{(N+2)\Pi g}{6(4\pi)^{2}} - \frac{5(N+2)\Pi^{2}g^{2}}{36(4\pi)^{4}}.
\end{eqnarray}
The eq. (\ref{gamma m}) shows us that its first and second terms, which are a result of one- and two-loop integrals, have extra factors of $\Pi$ and $\Pi^{2}$, respectively, when compared with its LI counterpart. A similar behavior was shown for the $\beta$ and $\gamma$ functions both, using explicit calculations, up to two-loop order and, by a redefinition of coordinates at least at lowest order in $K_{\mu\nu}$, up to any loop level, if we know the respective higher loop results for the LI theory \citep{PhysRevD.84.065030}. With the help of this redefinition, was possible to remove the $K_{\mu\nu}$ constants from the initial LV Lagrangian. As a consequence, a new Lagrangian (now a function of different coordinates) with the same form as for the LI theory with scaled parameters was obtained, namely the field and coupling constant. In particular, the new renormalized dimensionless coupling constant was shown to be $g \rightarrow \Pi g$. This explains why we have the behavior shown in the eq. (\ref{gamma m}). Thus we can use similar arguments as the ones used for obtaining the higher loop $\beta$ and $\gamma$ functions and write the $\gamma_{m}$ function for all loop order as
\begin{eqnarray}\label{gamma mn}
\gamma_{m}(g) = \sum_{n=1}^{\infty}\gamma_{m,n}^{(0)}\Pi^{n}g^{n}
\end{eqnarray}    
where $\gamma_{m,n}^{(0)}$ is the corresponding LI $n$th loop quantum correction to this function.

\section{Conclusions}\label{Conclusions}

\par We calculated the $\gamma_{m}$ function associated to mass renormalization for $O(N)$ self-interacting scalar field theory with Lorentz violation. For this purpose, this function was computed in MS where the Feynman diagrams were regularized using DR regularization in $d = 4 - \varepsilon$. As a result, we showed, by explicit calculations up to two-loop order and by coordinates redefinitions arguments up to any loop level, that the $\gamma_{m}$ function can be obtained of its LI version by a simple change in renormalized dimensionless coupling constant $g \rightarrow \Pi g$. This completes the task of getting a full renormalization of the theory and opens the possibility of studying a full possible LV extension of Higgs sector for the SM and further applications in scattering amplitude calculations.     

\appendix
\section{Integral formulas in $d$-dimensional euclidean momentum space}\label{Integral formulas in $d$-dimensional Euclidean momentum space}

\begin{eqnarray}
\int d^{d}q \frac{q^{\mu}}{(q^{2} + 2pq + M^{2})^{\alpha}} = -\hat{S}_{d}\frac{1}{2}\frac{\Gamma(d/2)}{\Gamma(\alpha)}\frac{p^{\mu}\Gamma(\alpha - d/2)}{(M^{2} - p^{2})^{\alpha - d/2}},
\end{eqnarray}

\begin{eqnarray}
\int d^{d}q \frac{q^{\mu}q^{\nu}}{(q^{2} + 2pq + M^{2})^{\alpha}} = \hat{S}_{d}\frac{1}{2}\frac{\Gamma(d/2)}{\Gamma(\alpha)}\left[ \frac{1}{2}\delta^{\mu\nu}\frac{\Gamma(\alpha - 1 - d/2)}{(M^{2} - p^{2})^{\alpha - 1 - d/2}} + p^{\mu}p^{\nu}\frac{\Gamma(\alpha - d/2)}{(M^{2} - p^{2})^{\alpha - d/2}} \right],
\end{eqnarray}

\begin{eqnarray}
\int d^{d}q \frac{q^{\mu}q^{\nu}q^{\rho}}{(q^{2} + 2pq + M^{2})^{\alpha}} = - \hat{S}_{d}\frac{1}{2}\frac{\Gamma(d/2)}{\Gamma(\alpha)} \left[\frac{1}{2}[\delta^{\mu\nu}p^{\rho} + \delta^{\mu\rho}p^{\nu} + \delta^{\nu\rho}p^{\mu}]\frac{\Gamma(\alpha - 1 - d/2)}{(M^{2} - p^{2})^{\alpha - 1 - d/2}} + \right.  \nonumber \\  \left. p^{\mu}p^{\nu}p^{\rho}\frac{\Gamma(\alpha - d/2)}{(M^{2} - p^{2})^{\alpha - d/2}} \right],
\end{eqnarray}

\begin{eqnarray}
\int d^{d}q \frac{q^{\mu}q^{\nu}q^{\rho}q^{\sigma}}{(q^{2} + 2pq + M^{2})^{\alpha}} = \hat{S}_{d}\frac{1}{2}\frac{\Gamma(d/2)}{\Gamma(\alpha)}\left[\frac{1}{4}[\delta^{\mu\nu}\delta^{\rho\sigma} + \delta^{\mu\rho}\delta^{\nu\sigma} +\delta^{\mu\sigma}\delta^{\nu\rho}] \frac{\Gamma(\alpha - 2 - d/2)}{(M^{2} - p^{2})^{\alpha - 2 - d/2}} + \right.  \nonumber \\  \left. \frac{1}{2}[\delta^{\mu\nu}p^{\rho}p^{\sigma} + \delta^{\mu\rho}p^{\nu}p^{\sigma} + \delta^{\mu\sigma}p^{\nu}p^{\rho} + \delta^{\nu\rho}p^{\mu}p^{\sigma} +\delta^{\nu\sigma}p^{\mu}p^{\rho} +\delta^{\rho\sigma}p^{\mu}p^{\nu}]\frac{\Gamma(\alpha - 1 - d/2)}{(M^{2} - p^{2})^{\alpha - 1 - d/2}}  + \right.  \nonumber \\  \left. \times p^{\mu}p^{\nu}p^{\rho}p^{\sigma}\frac{\Gamma(\alpha - d/2)}{(M^{2} - p^{2})^{\alpha - d/2}} \right].
\end{eqnarray}

\bibliography{apstemplate}

\end{document}